\documentclass{article}
\usepackage{graphicx}

 \usepackage[dblblindworkshop, final]{neurips_2025}

\usepackage[utf8]{inputenc} 
\usepackage[T1]{fontenc}    
\usepackage{hyperref}       
\usepackage{url}            
\usepackage{booktabs}       
\usepackage{amsfonts}       
\usepackage{amssymb}
\usepackage{nicefrac}       
\usepackage{microtype}      
\usepackage{xcolor}         

\title{Carbox: an end-to-end differentiable astrochemical simulation framework}

  \author{%
Gijs Vermariën$^{*1,2}$ \quad Tommaso Grassi$^{3,4}$ \quad Marie Van de Sande$^{1}$ \\ 
\textbf{ \quad Serena Viti$^{1,5,6}$ \quad Stefano Bovino$^{7,8,9}$ \quad Alessandro Lupi$^{10}$} \\ \textbf{Alexander Ruf$^{11,4}$ \quad Lorenzo Branca$^{12}$ \quad Catherine Walsh$^{13}$} \\
$^1$Leiden Observatory \quad
$^2$SURF \quad
$^3$MPE \quad 
$^4$Origins \quad \\
$^5$University of Bonn \quad
$^6$UCL \quad
$^7$University of Rome Sapienza \\
$^8$Universidad de Concepci\'on \quad
$^9$INAF Arcetri \quad
$^{10}$Università degli Studi dell’Insubria \quad\\
$^{11}$LMU Munich \quad
$^{12}$Heidelberg University \quad
$^{13}$University of Leeds\\
\texttt{* vermarien@strw.leidenuniv.nl}
}

\begin{document}

\maketitle

\begin{abstract}
 Since the first observations of interstellar molecules, astrochemical simulations have been employed to model and understand its formation and destruction pathways. With the advent of high-resolution telescopes such as JWST and ALMA, the number of detected molecules has increased significantly, thereby creating a need for increasingly complex chemical reaction networks. To model such complex systems, we have developed Carbox, a new astrochemical simulation code that leverages the modern high-performance transformation framework Jax. 
With Jax enabling computational efficiency and differentiability, Carbox can easily utilize GPU acceleration, be used to study sensitivity and uncertainty, and interface with advances in Scientific Machine Learning.
All of these features are crucial for modeling the molecules observed by current and next-generation telescopes.
\end{abstract}

\section{Astrochemical reaction networks}
The practice of simulating molecules in space has been a key tool since their earliest observations. However, with increasing resolution in both space and frequency over the entire electromagnetic spectrum, the number of molecular detections has risen sharply to 340\footnote{https://cdms.astro.uni-koeln.de/classic/molecules}. Likewise, the modeling of reaction networks has evolved from relatively simple hydrogen-based primordial (early universe) chemistry to increasingly complex organic molecular species that are the precursors to biological life. Many codes that evolve the corresponding differential equations to model the abundances of chemical species as a function of time and space have been developed in FORTRAN and C, with some exceptions in Python. The most notable are UCLCHEM\citep{holdshipChemulatorFastAccurate2021}, Nautilus\citep{ruaudGasGrainChemical2016}, and Magickal\citep{garrodFormationComplexOrganic2022a}, which focus solely on the chemical evolution, and KROME\citep{grassiKROMEPackageEmbed2014}, PRIZMO\citep{grassiModellingThermochemicalProcesses2020a}, and Grackle\citep{smithGRACKLEChemistryCooling2017a}, which also consider time-dependent coupled thermal processes. Although these codes are applied to different environments and chemical conditions, and may contain various chemical networks that differ in terms of size and composition, they are all based on standard implicit numerical ordinary differential equation (ODE) solvers.

The primary numerical integration method is the Backward Differentiation Formula (BDF) \citep{byrnePolyalgorithmNumericalSolution1975b}, which is well-suited for solving (very) stiff problems, i.e., the ODEs that dictate the chemistry. Apart from some outliers, this integration method dominated the astrochemical codes, with some variants, from the initial implementations (GEAR \cite{hindmarshGEAROrdinaryDifferential1972}) to modern SUNDIALS \citep{byrne_sensitivity_2024}. These methods have improved by exploiting numerical efficiency, for example, DLSODES, which takes advantage of the Jacobian sparsity of the chemical ODEs, or simply by leveraging the capabilities of newer and improved CPUs.

Conversely, the few examples of GPU-based solvers are subdominant and limited by memory, by well-established computational techniques available to CPU users, or by the limited coding capabilities and human resources within the astrochemical community.

Recently, scientific machine learning (SciML) approaches to evolve astrochemical systems have been explored by several groups \citep{grassiReducingComplexityChemical2021,holdshipChemulatorFastAccurate2021,brancaNeuralNetworksSolving2022,tangReducedOrderModel2022,sulzerSpeedingAstrochemicalReaction2023,brancaEmulatingInterstellarMedium2024,maesMACEMachineLearning2024,vermarienNeuralPDRNeuralDifferential2025}, however, the stiffness, the aforementioned non-linearity, and the high dimensionality of chemical systems have limited the applicability of these methods to realistic observed environments.
In fact, a more effective use of (SciML) has been the coupling of classical time-integration codes (or methods) to statistical inference pipelines \citep{holdship_energizing_2022,heyl_identifying_2022,behrens_neural_2024}, used to analyze the sensitivity of both chemical and physical parameters \cite{heyl_statistical_2023,vermarien_understanding_2025,grassiMappingSyntheticObservations2025}. This approach enables us to utilize statistical methods to predict the input parameters of chemical systems, including their physical boundary conditions. These methods try to determine how the output of a chemical model is influenced by any input parameter, either chemical or physical, sometimes trying to minimize the output in relation to the corresponding observable counterpart. Fully differentiable modeling can help both with the forward sensitivity analysis and enable better inverse parameter studies. 

Thanks to the Jax framework \citep{jax2018github}, Carbox\footnote{https://github.com/gijsvermarien/carbox} is a fully differentiable open-source astrochemical simulation framework that can simulate the evolution of chemical species. By utilizing the power of Jax, Carbox has native support for GPU acceleration and numerical automatic differentiation. Carbox is thus interoperable with SciML within the Jax ecosystem, allowing easier development of Machine Learning (ML) methods and even classical-ML hybrid solutions. It also means that Carbox can seamlessly be integrated with other emerging astrophysical codes implemented in Jax, such as radiative transfer and hydrodynamics \citep{storcks_differentiable_2024}. 

\section{Simulating astrochemical reaction networks}
A chemical kinetic reaction network consists of a set of chemical species connected by a set of reactions. In astrochemistry, the reactions, their reaction mechanism, and the reaction rate coefficients ($k_j$) can be inferred from theoretical studies, observations, or laboratory measurements.  The mathematical counterpart of a chemical network is a set of differential equations that describes the evolution in the number density $n_i(t)$ in $\mathrm{cm}^{-3}$ of each chemical species. The numerical problem becomes to solve an initial value problem (IVP), where, if we omit reactions with order greater than two, we have
$$ \frac{\mathrm{d}n_i}{\mathrm{d}t}=\pm\sum\nolimits_j k_jn_{s_1\in j}\pm\sum\nolimits_j k_j n_{s_1\in j}n_{s_2\in j}\,, $$
where the subscripts $s_1\in j$ and $s_2\in j$ indicate respectively the first and the second product or reactant involved in the $j$th reaction; both the unimolecular and bimolecular reactions lead to the production (+) and destruction (-) of the species $i$. The ODE system is integrated in time, given some initial conditions $n_i(t=0)$ for each $i$th species.

There are several classes of rate reaction coefficients, and their exact expression depends on the chemistry involved \citep{bovino_astrochemical_2023}. Their description is beyond the aims of this paper; nevertheless, to illustrate the nature of our systems, we provide two notable examples.
For a two-body reaction in the gas phase, the reaction rate is described in general by a temperature-dependent function, and often by the Kooji-Arrhenius equation,
$$
k = \alpha{T_{300}}^\beta \exp(-\gamma/T)\,,
$$with $\alpha$, $\beta$, and $\gamma$ specific reaction parameters, and $T_{300}$ the temperature in units of 300\,K. On the other hand, reaction coefficients can be functions of parameters other than temperature, depending on the specific core process or environmental conditions, as ultraviolet and X-ray photons, and cosmic rays, which ionize or dissociate species with a rate that usually depends on their energy source. For example, for the direct cosmic-rays ionization (${{\rm X}+{\rm CR}\to {\rm X}^++{\rm e}^-}$) we have  $$k = \alpha\,\zeta\,,$$
with $\alpha$ a parameter depending on the given chemical species (X), and $\zeta$ the ionization rate per H$_2$ molecule, a value that depends on the modeled astrophysical environment, and which plays a crucial role in the evolution of some observed chemical species.

This suggests that the ODE system can be generalized as $\dot{n}(t)=f(n(t), k(\varphi), \varphi)$, where $n$ are all the chemical abundances, $k$ the rate coefficients, and $\varphi$  the environment parameters (e.g., $T$, $\zeta$ or $F_\mathrm{UV}$). This highlights the strong dependence of astrochemical systems on both the physical conditions and the reaction networks, as well as their strong nonlinearity when evolved over time.

\section{Sensitivity analysis of astrochemical models}

Astrochemical models are often used in a deterministic manner, however, in many realistic applications, it is necessary to determine the effect of varying these parameters.
In addition, several of these parameters have intrinsic (large) uncertainties that play a key role in shaping the scientific interpretation of the results.
This includes the aforementioned physical input parameters, as well as the rate coefficients $k$.
Assessing the effect of uncertainties in the model input on the model predictions has been done for a variety of astrochemical environments, spanning all stages of stellar and galactic evolution \citep{wakelam_estimation_2005,vasyunin_chemistry_2008,dobrijevic_comparison_2010,penteado_sensitivity_2017,maes_sensitivity_2023}.

The uncertainties on the rate coefficients, i.e., the kinetic data that drive the chemical evolution, can lead to large errors in the abundances predicted by the model.
Fig. \ref{fig:crichoutflow} shows the predicted abundances of a set of molecules throughout a stellar outflow, which is a relatively standard astrochemical benchmark. 
For some species, the uncertainty in the predicted abundances is limited. For others, the uncertainty spans several orders of magnitude. 
This has important implications for the interpretation of astrophysical observations, as some species are more reliable probes of the physical conditions than others, and relatively minor discrepancies might lead to different physical interpretations.
It also has repercussions for astrochemical model development, as adding chemical or physical complexity might not yield significantly different results, hence over-complicating the model \citep{de_sande_sense_2025}.
Finally, determining which reactions' uncertainties have the largest impact on the predicted chemistry is a crucial step in planning costly computational quantum chemistry calculations and laboratory experiments.

A modern and effective approach to these problems is to use differentiable models, which enable us to more directly assess both the forward and inverse problems. One of the main aims of Carbox is to be a differentiable tool that enables uncertainty and sensitivity analysis with respect to any of the input parameters, both physical and chemical.

\section{Carbox}
Carbox aims to provide a modern open-source implementation of an astrochemical simulation framework, with a core routine that can build a system of differential equations based on the species and reactions in a given arbitrary astrochemical network. The core routine relies on these species and reactions to have a uniform format, also providing the capability to parse three existing simulation frameworks: PRIZMO\citep{grassiModellingThermochemicalProcesses2020a}, UMIST\citep{millarUMISTDatabaseAstrochemistry2024}, and UCLCHEM\citep{holdship_uclchem_2017}. These parsers can be easily extended to different reaction database formats and custom rate equations. After parsing the species and reactions and providing them to Carbox, the framework constructs the incidence matrix to enable the effective time integration of the corresponding ODE system. The result of the parsing is a \emph{Network} object that provides easy user interaction and a mirrored \emph{JNetwork} object that represents the mathematical evolution of the chemical network and is compiled by and interoperable with Jax.
Support is added for the vectorization of rates with identical mechanisms, as well as for sparsity, providing a tradeoff between memory footprint and
computational speed, which is needed for large ($\gtrsim 1000$ reactions) chemical networks. These networks can then
be integrated using Diffrax\citep{kidgerNeuralDifferentialEquations2022},
specifically, with the 5th order Kvaerno stiff integrator and custom tolerances of $a_{tol}=10^{-9}$ and $r_{tol}=10^{-30}$ providing reliable results \citep{kvaerno_singly_2004}.

\begin{figure}
    \centering
    \includegraphics[width=1\linewidth]{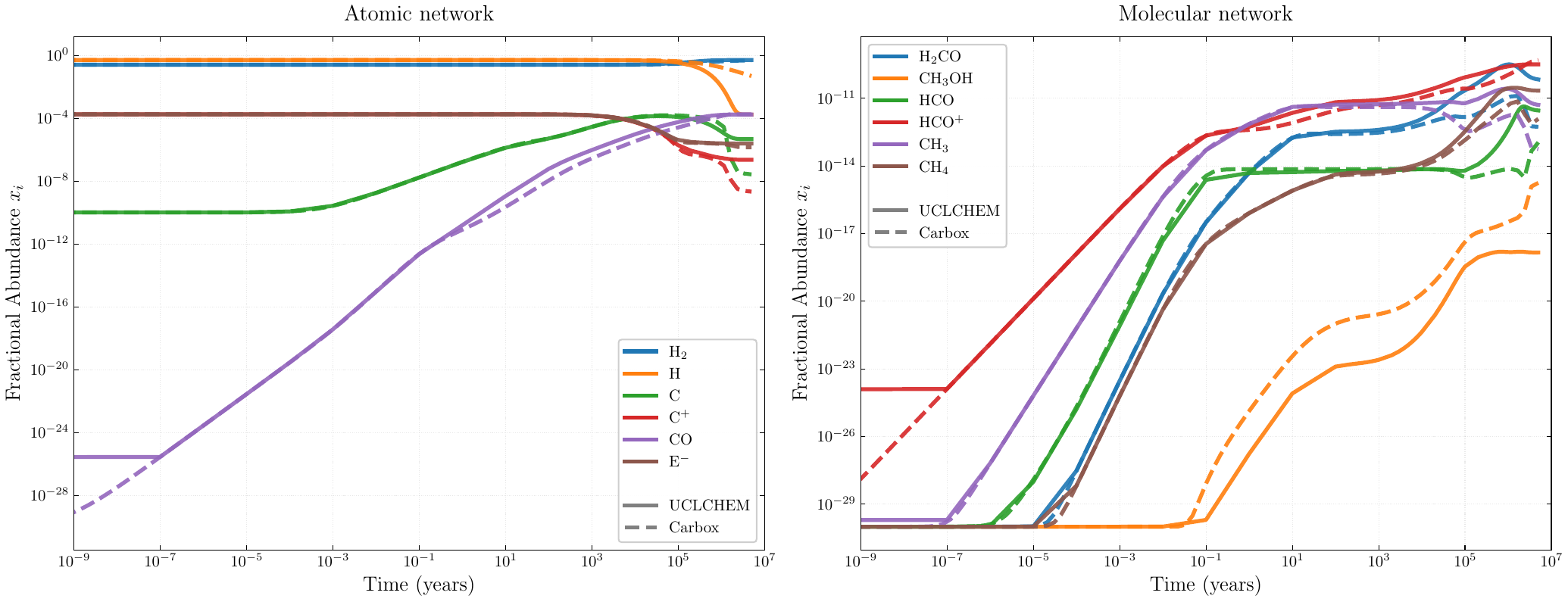}
    \vspace{-6mm}
     \caption{\small Temporal evolution of the fractional abundances of some selected chemical species for the atomic (left) and molecular (right) reaction networks. The solid line represents the reference integration using the UCLCHEM gas-phase reaction network, while the dashed lines show the results obtained with Carbox. The results are similar for the species not displayed here.}
    \label{fig:evolution}
\end{figure}

To verify the robustness of Carbox, we benchmark it against a gas-phase-only setup of UCLCHEM, using two different networks: a relatively small minimal network with only atomic species and a more extensive network that includes molecules such as methanol (CH$_3$OH). Given the stiffness of the corresponding ODEs, we can generalize the accuracy of the results to more complicated chemical networks. The resulting simulation can be seen in Fig. \ref{fig:evolution}. Showing we can reliably simulate the reaction network. It also highlights the non-linear nature of astrochemical models; small implementation differences can result in order-of-magnitude differences. However, since the ice chemistry is not included, the model provides a feasible approximation of the imperfect gas-phase only abundances.

\begin{figure}
    \centering
    \includegraphics[width=1\linewidth]{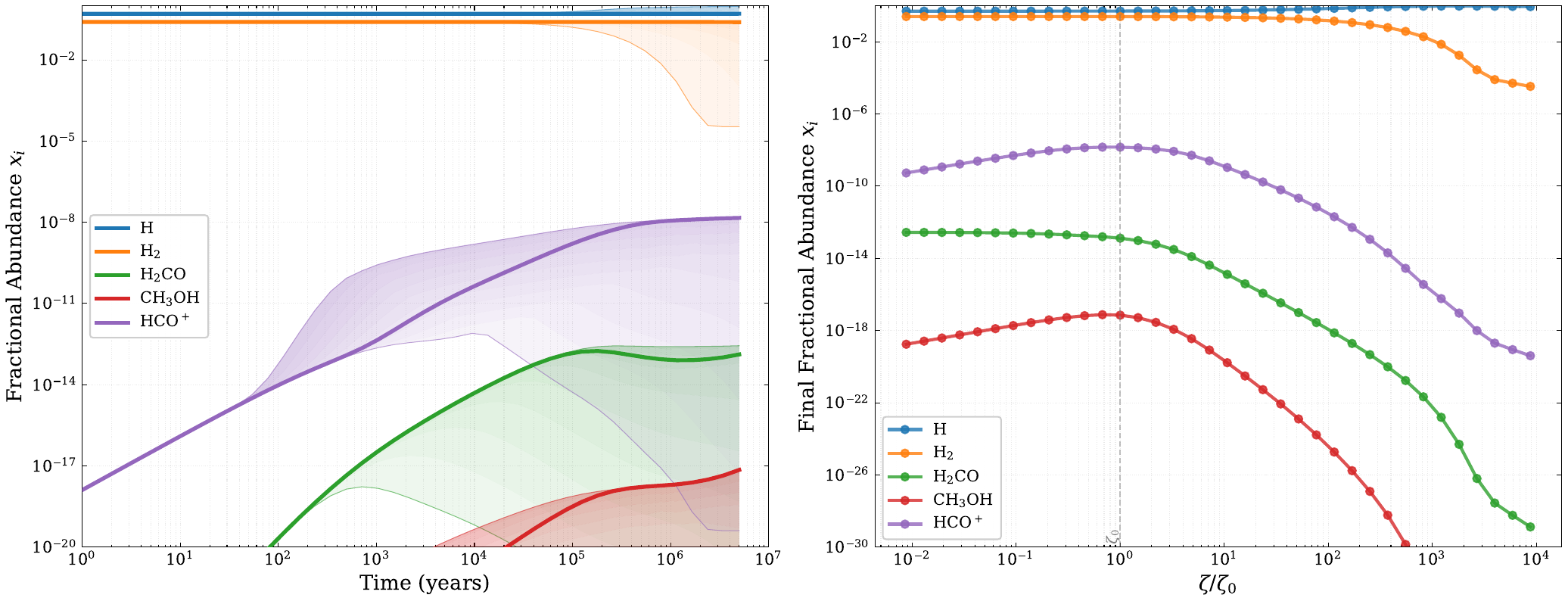}
    \vspace{-6mm}
    \caption{\small Left: temporal evolution of selected chemical species. The solid line represents the reference evolution ($\zeta/\zeta_0=1$), while the shaded area is the maximum extent given by $\zeta/\zeta_0\in[10^{-2}, 10^4]$. Right: the final ($t=10^7$\,yr) abundances of the same species for different cosmic-ray ionization rate values. }
    \label{fig:placeholder}
\end{figure}

One of the goals of Carbox is to assist in sensitivity studies. We test the impact of varying the cosmic ray ionization rate $\zeta$, denoted in units of the galactic cosmic ray ionization rate $\zeta_0=1.3\cdot10^{-17}\;\mathrm{s^{-1}}$, on the evolution of different molecular species. This parameter is present in a subset of only 20 out of 2227 reactions involving a subset of chemical species. As expected, varying the ionization efficiency determines a considerable variation in the temporal evolution (Fig.\,\ref{fig:placeholder}, left panel) and in the final ($t=10^7$\,yr) abundances (right panel), both in the directly-affected chemical species (e.g. H$_2$), but also on the species that are modified indirectly in a non-linear fashion (e.g., H$_2$CO), being the latter usually difficult to predict even in relatively simple chemical networks.

\section{Conclusion}
We introduced the open-source Carbox simulation framework, a differentiable Jax-based framework to integrate the ODE system corresponding to astrochemical networks. Our code reproduces the results of well-established astrochemical codes, but it enables differentiability and GPU acceleration; a crucial feature that enables computationally efficiently solving forward and inverse problems.

\newpage
\section*{Acknowledgments}
The authors declare no competing interests. This work is financially supported by the advanced ERC grant ( ID: 833460, PI: Serena Viti) within the framework of MOlecules as Probes of the Physics of EXternal (MOPPEX) project. C.W.~acknowledges financial support from the Science and Technology
Facilities Council and UK Research and Innovation (grant numbers
ST/X001016/1 and MR/T040726/1).
MVdS thanks the Oort Fellowship at Leiden Observatory. This research was supported by the Munich Institute for Astro-, Particle and BioPhysics (MIAPbP) which is funded by the Deutsche Forschungsgemeinschaft (DFG, German Research Foundation) under Germany´s Excellence Strategy – EXC-2094 – 390783311. SB acknowledges BASAL Centro de Astrofisica y Tecnologias Afines (CATA), project number AFB-17002.

\section*{Full affiliations}
{
\small
\begin{enumerate}
\item Leiden Observatory, Leiden University, P.O. Box 9513, 2300 RA Leiden, The Netherlands
\item SURF, Amsterdam, The Netherlands
\item Center for Astrochemical Studies, Max Planck Institut f\"ur Extraterrestrische Physik, Garching, Germany
\item Exzellenzcluster Origins, Boltzmannstr. 2, 85748 Garching, Germany
\item Transdisciplinary Research Area (TRA) ‘Matter’/Argelander-Institut für Astronomie, University of Bonn, Bonn, Germany
\item Department of Physics and Astronomy, University College London, Gower Street, London, UK
\item Department of Chemistry, University of Rome Sapienza, P.le A. Moro 5, 00185 Rome, Italy
\item Departamento de Astronom\'ia, Facultad de Ciencias F\'isicas y Matem\'aticas, Universidad de Concepci\'on, Av. Esteban Iturra s/n Barrio Universitario, Casilla 160, Concepci\'on, Chile
\item INAF, Osservatorio Astrofisico di Arcetri, Largo E. Fermi 5, I-50125 Firenze, Italy
\item Como Lake Center for Astrophysics, DiSAT, Università degli Studi dell’Insubria, via Valleggio 11, 22100 Como, Italy
\item LMU Munich, Faculty of Physics, Schellingstraße 4, 80799 Munich, Germany
\item Interdisciplinary Center for Scientific Computing, Heidelberg University, Heidelberg, Germany
\item School of Physics and Astronomy, University of Leeds, Leeds LS2 9JT, UK
\end{enumerate}
}

{
\small
\bibliographystyle{plainnat}
\interlinepenalty=10000
\bibliography{references.bib}
}

\appendix
\section{Sensitivity analysis of reaction rates}
\begin{figure}
    \centering
    \includegraphics[width=0.5\linewidth]{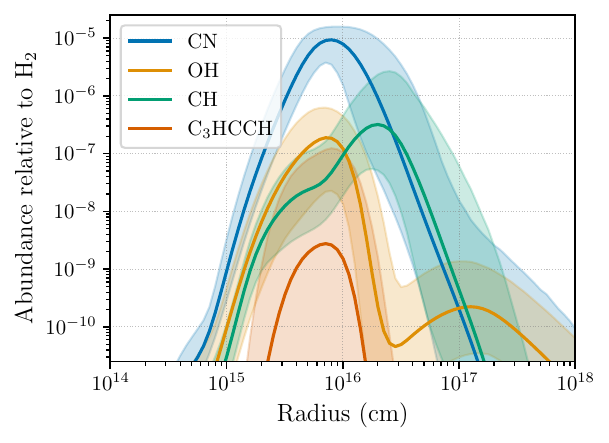}
    \caption{A figure illustrating the sensitivity of a carbon rich stellar outflow with a MCMC sampler varying reaction rates of the entire network. }
    \label{fig:crichoutflow}
\end{figure}

In figure \ref{fig:crichoutflow}, we illustrate a sensitivity analysis of a stellar outflow. We can see that molecules such as CN have a relatively tight peak in abundances, with low sensitivity to the reaction rates, whereas larger carbon molecules such as C$_3$HCCH have a large sensitivity.

\section{Reaction networks}
\subsection{Atomic Network}
The atomic networks describes the most basic of reaction networks containing only 31 species and 380 reactions:
$\mathrm{
C,\ C^{+},\ CH,\ CH^{+},\ CH_2,\ CH_2^{+},\ CH_3,\ CH_3^{+},\ CH_4,\ CH_4^{+},\ CH_5^{+},\ CO,\ CO^{+},\ e^{-},}$\\
$\mathrm{H,\ H^{+},\ H_2,\ H_2^{+},\ H_2O,\ H_2O^{+},\ H_3^{+},\ H_3O^{+},\ HCO,\ HCO^{+},\ He,\ He^{+},\ Mg,\ Mg^{+}, O,\ O^{+},\ O_2,}$\\
$\mathrm{
 \ O_2^{+},\ OH,\ OH^{+}
}$. The reactions contain cosmic ray ionization reactions, photon reactions with rate:
$$ 
k=\alpha F_{\mathrm{UV}} \exp(-\mathrm{k}A_\mathrm{V})
$$
where $F_\mathrm{UV}$ is the strength of the radiation field often in Habing units, and $A_\mathrm{v}$ the visual extinction, which is a measure of the attenuation of the radiation due to the presence of interstellar dust grains, and it depends on the reactant and on the dust optical properties. They contain reactions via secondary photons
induced by cosmic rays: 
$$ \alpha(T_{300})^\beta\frac{E}{1-\omega}\zeta
$$
where $E$ is the cosmic ray efficiency and $\omega$ is the dust grain albedo. This is then combined with
all the reactions from UMIST that contain all the species in the network. Lastly, we use a special H$_2$ formation 
and dissociation, CO dissociation and C$^+$ ionization routine that is implemented like-for-like from UCLCHEM.

\subsection{Molecular network}
The molecular network expands on the atomic network by including the same reaction mechanisms, but now with
more species. These species would normally be allowed to freeze out onto dust particles at lower temperatures, introducing ice chemistry that allows larger complex organic molecules such as methanol to form more effectively
\cite{jimenez-serra_modelling_2025}. This is turned off however, since the ice-chemistry is not yet implemented in 
Carbox. This results in a total of 161 species with 2227 reactions. 

\end{document}